\documentclass[ aip, jmp, amsmath,amssymb, reprint, ]{revtex4-1}
\usepackage{graphicx}% Include figure files
\usepackage{dcolumn}% Align table columns on decimal point
\usepackage{bm}% bold math
\usepackage{amssymb}
\usepackage{amsmath}
\usepackage{amsfonts}
\usepackage{comment}

\usepackage{hyperref}
\usepackage{bbm}
\usepackage[usenames]{color}
\begin{document}

\preprint{AIP/123-QED}

\title{Travelling waves in an ensemble of excitable oscillators: the interplay of memristive coupling and noise}
%\title{Memristive coupling supports travelling waves in an ensemble of excitable oscillators: deterministic and stochastic dynamics} % Force line breaks with \\
%\thanks{Footnote to title of article.}

\author{Ivan A. Korneev}
\affiliation{Institute of Physics, Saratov State University, Astrakhanskaya str. 83, 410012 Saratov, Russia}

\author{Ibadulla R. Ramazanov}
\affiliation{Institute of Physics, Saratov State University, Astrakhanskaya str. 83, 410012 Saratov, Russia}

\author{Andrei V. Slepnev}
\affiliation{Institute of Physics, Saratov State University, Astrakhanskaya str. 83, 410012 Saratov, Russia}

\author{Tatiana E. Vadivasova}
\affiliation{Institute of Physics, Saratov State University, Astrakhanskaya str. 83, 410012 Saratov, Russia}

\author{Vladimir V. Semenov}
\email{semenov.v.v.ssu@gmail.com}
\affiliation{Institute of Physics, Saratov State University, Astrakhanskaya str. 83, 410012 Saratov, Russia}

\date{\today}% It is always \today, today,
            
\begin{abstract}
Using methods of numerical simulation, we demonstrate the constructive role of memristive coupling in the context of the travelling wave formation and robustness in an ensemble of excitable oscillators described by the FitzHugh-Nagumo neuron model. First, the revealed aspects of the memristive coupling action are shown on an example of the deterministic model where the memristive properties of the coupling elements provide for achieving travelling waves at lower coupling strength as compared to non-adaptive diffusive coupling. In the presence of noise, the positive role of memristive coupling is manifested as significant increasing a noise intensity critical value corresponding to the noise-induced destruction of travelling waves as compared to classical diffusive interaction. In addition, we point out the second constructive factor, the L{\'e}vy noise whose properties provide for inducing travelling waves.
\end{abstract}

%Use showkeys class option if keyword
\maketitle

\begin{quotation}
Travelling waves of different nature and properties are observed in many physical, biological and chemical media and networks. Models exhibiting such structures are widely used for description of a broad spectrum of regular and chaotic spatio-temporal evolutional processes from signal and energy transmission to climate change, spread of epidemics, etc. To induce and suppress travelling waves and to control their characteristics, one can vary the model parameters as well as apply regular or stochastic external forcing. One more approach for controlling travelling waves can be realized in networks of coupled oscillators: a control scheme based on tuning the coupling properties (for instance, the coupling strength, topology and adaptivity). In the current paper, we address this issue in the context of interplay of adaptive coupling and noise perturbations. In addition, we extend a manifold of noise-induced effects observed in high-dimensional systems by L{\'e}vy-noise-induced travelling waves characterized by higher robustness in the presence of  memristive coupling. 
\end{quotation}

\section{Introduction}
\label{intro}
Travelling waves represent an interdisciplinary phenomenon uniting an incredibly broad variety of dynamical processes in deterministic and stochastic media \cite{landa1996,ghazaryan2022,smoller1994,garcia1999,shepelev2016}, networks \cite{korneev2021,korneev2022,semenov2023,zakharova2023,rybalova2023} and delayed-feedback oscillators \cite{giacomelli2012,klinshov2017,semenov2018,zakharova2024}. Exhibited as spatially-periodic structures, fronts, backs and pulses, travelling waves are of a frequent occurrence in plasma physics \cite{loarte1999,schwabe2007}, optics and electronics \cite{ebeling1993,akhmediev1997,semenov2023-2,liu2016}, hydrodynamics \cite{henry2019,debnath1994,johnson1997,dias1999}, chemistry \cite{kuramoto1984,kapral1995,epstein1998}, neurophysiology \cite{hugles1995,muller2018,mohan2024}, as well as on the edge of ecology, population biology and epidemiology \cite{sherratt2008,malchow2008,brauer2012}. Such diversity causes  interest of specialists in nonlinear dynamics and complex systems focused on revealing the interdisciplinary, fundamental properties of travelling waves and controlling their characteristics and stability.

In the current paper, travelling waves are studied in an ensemble of stochastic excitable oscillators with local adaptive coupling. The features of coupling are associated with the properties of the coupling element, the memristor. The memristors have a wide range of practical applications, first of all, neuromorphic computing and the implementation of new generation memory elements \cite{kozma2012,chua2022}. In the context of nonlinear dynamics, the memristor is interesting as an element whose intrinsic properties can essentially change the dynamics of oscillatory systems and are responsible for qualitatively new types of the behavior from bifurcations without parameters in single memristor-based oscillators with lines of equilibria \cite{korneev2017,korneev2017-2,korneev2021-2,korneev2023,korneev2023-2} to the Turing patterns \cite{buscarino2016} and travelling waves in networks of memristive elements \cite{pham2012}. The presence of the memristor as a coupling element provides for the observation of initial-condition-dependent synchronization of regular \cite{korneev2020} and chaotic \cite{korneev2021-3} self-oscillators and wave processes in single-layer \cite{korneev2021} and multilayer \cite{korneev2022} networks. Based on the results presented in Ref. \cite{korneev2021} addressing the impact of memristive coupling on travelling waves in a deterministic network, we expand a spectrum of phenomena exhibited by complex systems and associated with the memristor action by the consideration of travelling waves being more and more robust against the stochastic impact when the memristive properties of coupling become more pronounced.

The repeatedly pointed out similarity between the memristor behaviour and the functional peculiarities of neural cell synapses  (for instance, see Refs. \cite{jo2010,li2013,serb2016,williamson2013}) has inspired us to consider an ensemble of excitable neurons coupled through the memristive coupling as a promising model for simulation of effects in biological neural networks. Noise sources with various characteristics inevitably present in biological neural networks and significantly affect the neuron dynamics \cite{lindner2004,pisarchik2023}. One of the most common models to describe fluctuations in such systems is  white or coloured Gaussian noise. In certain cases (for example, in the presence of abrupt stochastic impulses), stochastic processes with the L{\'e}vy distribution can model the dynamics of real biological neurons more accurately as compared to Gaussian noise \cite{nurzaman2011,wu2017}. Motivated by the significance of L{\'e}vy processes in neural systems, we consider a network of excitable elements subject to the L{\'e}vy white noise. Thus, our research is focused on two factors affecting travelling waves: the presence of memristive coupling and noise including occasional high-amplitude impulses. This choice is dictated by an attempt to describe effects that can be potentially observed in biological neural networks and to reveal fundamental peculiarities of travelling waves in complex systems with adaptive coupling.

\section{Model and methods}
\label{model_and_methods}
\begin{figure}[t]
\centering
\includegraphics[width=0.48\textwidth]{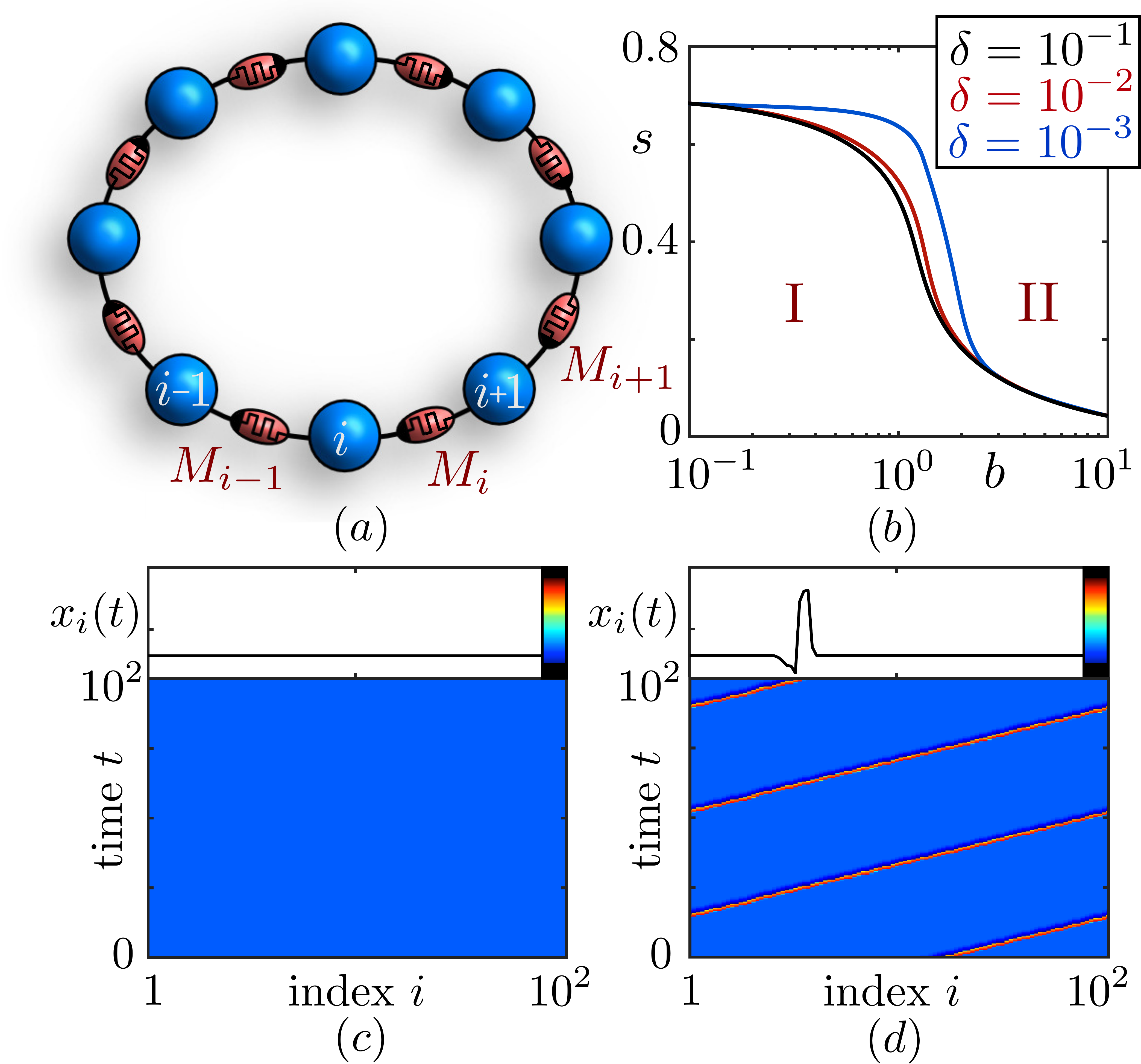}
\caption{(a) Schematic representation of an ensemble of the memristively coupled (through elements $M_i$) FitzHugh-Nagumo neurons (see Eqs. (\ref{eq:system})); (b) Map of regimes on the plane ($b$,$s$) characterising the deterministic system, $\xi_i(t)\equiv0$. The only regime of the collective dynamics in area I is the quiescent steady state regime (panel (c)). This regime coexists with travelling waves (particular wave obtained for $b=3$, $s=0.5$, $\delta=10^{-3}$ is depicted in panel (d)) in area II; (c)-(d) Space-time plots $x_i(t)$ illustrating the ensemble dynamics. The upper insets show the ensemble state at the last moment $t = 10^{2}$. The system parameter is $\varepsilon=0.01$.}
\label{fig1}
\end{figure}

The model under study is schematically illustrated in Fig.~\ref{fig1}~(a). It represents an ensemble of the identical FitzHugh-Nagumo oscillators in the excitable regime. The oscillators interact through local memristive coupling. The model equations are:
\begin{equation}
\label{eq:system}
\left\lbrace
\begin{array}{l}
\dfrac{dx_i}{dt}=\dfrac{1}{\varepsilon}\left(x_i-y_i - x^3_i/3\right)\\
+s \left[M(z_{i-1})(x_{i-1}-x_i)+ M(z_i)(x_{i+1} - x_i)\right],\\
\dfrac{dy_i}{dt}=0.8 x_i - y_i+0.2 +\xi_i(t) ,\\
\dfrac{dz_i}{dt}=x_i-x_{i+1}-\delta z_i,\\
\end{array}
\right.
\end{equation}
where $x_i$ and $y_i$ are the fast and slow dynamic variables which define the instantaneous state of the $i$-th oscillator ($i=1,2,\ldots, N$, where $N$ is the number of interacting oscillators). The parameter $\varepsilon$ is usually assumed to be small, which corresponds to the relaxation behaviour of the single oscillator (in the considered model, parameter $\varepsilon$ is chosen to be $\varepsilon=0.01$). Variables $z_i$ determine the instantaneous states of the memristive coupling elements, whose conductivity is given by the function $M(z_i)=1+bz_i^2$ (the cubic memristor model where $a$ and $b$ are parameters). The memristor state equations $\dot{z}_i=x_i-x_{i+1}-\delta z_i$ contain the parameter $\delta$ which characterizes the memristor forgetting effect. The larger is the parameter $\delta$, the shorter is the time range where the correlation between the initial and instantaneous states persists \cite{chang2011,chen2013,zhou2019}. To avoid the effects reported in paper \cite{korneev2021} and associated with a continuous dependence of the oscillatory dynamics characteristics on the initial conditions, case $\delta=0$ (the ideal memristor case) is excluded from the consideration. 

In the absence of noise, model (\ref{eq:system}) was considered in Ref. \cite{korneev2021} where the aspects of memristive coupling are described in more detail. In the current paper, we study the coupled FitzHugh-Nagumo oscillators in the excitable regime subject to L{\'e}vy noise. Statistically independent additive L{\'e}vy noise sources $\xi_i(t)$ in Eqs. (\ref{eq:system}) are defined as the formal derivatives of the L{\'e}vy stable motion. L{\'e}vy noise is characterized by four parameters: a stability index $\alpha \in (0:2]$, a skewness (asymmetry) parameter $\beta\in [-1:1]$, a parameter $\mu$ (is assumed to be zero in the current research) being a mean value of the L{\'e}vy noise when $1\leq \alpha \leq 2$ and a scale parameter $\sigma$. Parameter $D=\sigma^{\alpha}$ is introduced as the noise intensity. 
%The characteristic function of noise sources $\xi_i(t)$ takes the form \cite{janicki1994,dybiec2006,dybiec2007}:
To generate L{\'e}vy noise signals $\xi_i$, the Janicki-Weron algorithm was applied for the fixed parameter values $\beta=\mu=0$ and a variable scale factor $1<\alpha \leq 2$. In such a case, the formula for random number generation takes the simplified form (see papers \cite{janicki1994,weron1995} for more details):
%\begin{widetext}
\begin{equation}
\label{eq:noise_generation} 
\xi_i=\sigma \times \dfrac{\sin(\alpha V_i)}{(\cos(V_i))^{1/\alpha}}\times \left( \dfrac{\cos(V_i(1-\alpha))}{W_i}\right)^{\dfrac{1-\alpha}{\alpha}}, 
\end{equation} 
%\end{widetext}
where $V_i$ are random variables uniformly distributed on $\left(-\dfrac{\pi}{2}:\dfrac{\pi}{2}\right)$, $W_i$ are exponential random sequences with mean 1 (variables $W$ and $V$ are statistically independent). A similar numerical procedure was used in Ref. \cite{korneev2024} addressing the issue of L{\'e}vy noise-induced coherence resonance in the single FitzHugh-Nagumo oscillator in the excitable regime. In case $\alpha=2$, signals $\xi_i(t)$ represent independent sources of white Gaussian noise. In case $\alpha<2$, the noise distribution is non-Gaussian and contains long heavy tails associated with random impulses of high amplitude.

% noise with zero mean value and the intensity $D=\sigma^2$.
%the variance being equal to $2\sigma^2$. In case $\alpha<2$ the distribution is non-Gaussian and the variance is infinite.  

%%%%%%%%%%%%%%%%%%%%%%%% FIG 1 %%%%%%%%%%%
%%%%%%%%%%%%%%%%%%%%%%%%%%%%%%%%%%%%%%

Numerical simulations were carried out by the integration of model equations  (\ref{eq:system}) using the Heun method \cite{mannella2002} with the time step $\Delta t=10^{-3}$ or smaller. The boundary conditions are chosen to be periodic: $x_{i\pm N} = x_i$, $y_{i\pm N} = y_i$ and $z_{i\pm N} = z_i$.
At each simulation step, new random values $\xi_i(t_0+n\Delta t)$ were generated according to formula (\ref{eq:noise_generation}). It is important to note that numerical modelling of equations including $\alpha$-stable stochastic process with finite time step implies the normalization of the noise term by $\Delta t^{1/\alpha}$ (see also Refs. \cite{xu2016,pavlyukevich2010}). For this reason, the generated values $\xi_i$ (see formula (\ref{eq:noise_generation})) were normalized by this term.

During the numerical modelling process, sufficiently long time of transient processes were discarded to observe the established natural and stationary dynamics. Further exploration was carried out by the analysis of plotted instantaneous spatial profiles and spatio-temporal diagrams. 

\section{Deterministic dynamics}
\label{det_dyn}
First, model (\ref{eq:system}) was considered in the absence of noise, $\xi_i(t)\equiv 0$ to reveal the impact of memristive properties of coupling on travelling waves in the deterministic system. In such a case, one can induce travelling waves in the system by starting simulation from $x_i(0)=\sin(2\pi i/N),~y_i(0)=\cos(2\pi i/N)$, if the coupling strength $s$ is sufficiently high. 
Then continuous decreasing parameter $s$ allows to reveal the area of travelling wave existence\footnote{We continuously vary the coupling strengths with certain parameter step such that the last state at the previous iteration is the initial condition for the next one.}. Result of applying such method for different values of the memristor parameter $b$ is depicted in Fig.~\ref{fig1}~(b) as a map of regimes on the plane ($b$,$s$) which contains two areas. The only kind of collective dynamics in area I is a completely quiescent steady state regime corresponding to the realization $x_i(t)\approx-1.076$ and $y_i(t)\approx-0.661$ (coordinates of a stable steady state in the phase space of the single oscillator). In contrast, the coexistence of the quiescent regime and travelling waves is realized in area II. As can be seen in Fig. \ref{fig1}~(b), increasing the parameter $b$ being responsible for manifestation of memristive properties allows to achieve travelling waves at lower values of the coupling strength. Changing the parameter $\delta$ being responsible for the memristor forgetting effect allows to shift the boundary between the areas of the existence and the absence of travelling waves, but the revealed effect persists: memristor-based adaptive coupling provides for achieving travelling waves at lower coupling strength. 

\section{Impact of noise}
\subsection{Gaussian noise}
%%%%%%%%%%%%%%%%%%%%%%%% FIG 2 %%%%%%%%%%%
\begin{figure}[t!]
\centering
\includegraphics[width=0.48\textwidth]{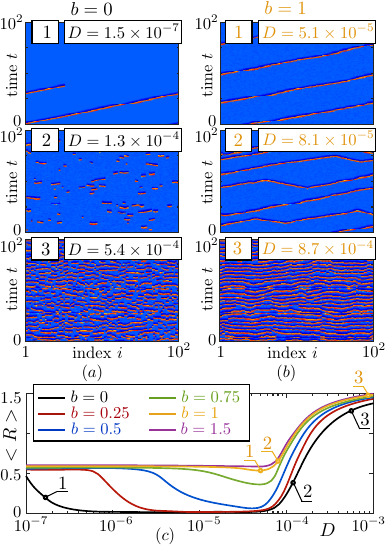}
\caption{Transformation of travelling waves when increasing the intensity $D$ of white Gaussian noise in the absence ($b=0$, panel (a)) and in the presence ($b=1$, panel (b)) of the coupling element memristive properties; (c) The dependencies of the characteristic $<R>$ on the noise intensity $D$ when increasing parameter $b$. Points 1-3 in curves for $b=0$ and $b=1$ correspond to space-time plots in panels (a) and (b). System parameters are $s=0.7$, $\varepsilon=0.01$, $\delta=10^{-2}$. The noise scale parameter is $\alpha=2$ (white Gaussian noise), the noise intensity is introduced as $D=\sigma^{\alpha}$.}
\label{fig2}
\end{figure}
%%%%%%%%%%%%%%%%%%%%%%%%%%%%%%%%%%%%%%
We begin studying the impact of fluctuations from the case of white Gaussian noise when fixing noise parameter $\alpha=2$ and varying $\sigma$. In the absence of memristive properties (at $b=0$), increasing the noise intensity results in suppression of travelling waves as demonstrated in Fig.~\ref{fig2}~(a1) and reduces the dynamics to the slightly fluctuating quiescent regime. Further increasing the noise level induces spontaneous local spiking activity which extends to the neighbour oscillators due to the action of coupling [Fig.~\ref{fig2}~(a2)]. When the noise intensity growth continues, spiking activity becomes more regular (the occurrence of coherence resonance) and spatially coherent (the manifestation of synchronization of noise-induced oscillations) as depicted in Fig.~\ref{fig2}~(a3). 

When the coupling elements exhibit the memristive properties ($b>0$), travelling waves become much more robust. To illustrate this fact, panel (b1) in Fig.~\ref{fig2} contains a travelling wave which persists at the noise intensity $D=5.1\times10^{-5}$. This value is 340 times higher than the critical value of the noise intensity corresponding to the noise-induced travelling wave destruction observed when the memristive properties of coupling are not expressed (compare Fig.~\ref{fig2}~(a1) and  Fig.~\ref{fig2}~(b1)). Moreover, as depicted in Fig.~\ref{fig2}~(b2), further growth of the noise intensity does not reduce travelling waves to the quiescent dynamics, but transforms this regime to wandering wave motion including the interaction of the initial wave with new ones spontaneously induced by noise. Noise of larger intensity induces synchronization of the noise-induced oscillations [Fig.~\ref{fig2}~(b3)]. Despite the space-time plot in Fig.~\ref{fig2}~(b3) is obtained at higher noise-intensity as compared to Fig.~\ref{fig2}~(a3), the effect of synchronization is more pronounced which indicates the constructive role of memristive coupling in the context of both travelling waves and the phenomenon of synchronization.

To quantitatively describe the impact of noise in the context of both travelling wave suppressing and inducing, we have introduced a characteristic $R$ involving the deviations of dynamical variable values $x_i(t)$ from the steady state coordinate $x_*\approx-1.076$ and calculated as being: $R=\dfrac{1}{N}\sum \limits_{i=1}^{N}\text{RMS}(x_i(t)-x_*)$, where $\text{RMS}$ means the root mean square. Starting from the initial conditions representing a single wave obtained in the deterministic model for chosen set of parameters (like depicted in Fig.~\ref{fig1}~(d)), we integrated stochastic equations (\ref{eq:system}) with the same parameter set except of noise parameters. Since the travelling waves are suppressed and induced unpredictably at different time moments, we integrated model (\ref{eq:system}) in time $t\in [0:200]$, repeated the numerical experiments 50 times and calculated the mean value $<R>$ describing the noise-induced effects as the stationary dynamics manifestation. 

Panel (c) in Fig. \ref{fig2} illustrates the dependencies of the quantity $<R>$ on the Gaussian noise intensity $D$ (the scale parameter is $\alpha=2$) for varying parameter $b$ being responsible for the memristive properties of the coupling elements. The black curve in Fig.~\ref{fig2}~(c)  characterising the ensemble at $b=0$ possesses a sharp drop at $D\in [10^{-7}:3\times 10^{-7}]$ corresponding to more and more quick noise-induced travelling wave suppression and achieving the quiescent steady state regime. Further noise intensity growth results in noise-induced spiking activity which gives rise to increasing $<R>$. As demonstrated in Fig.~\ref{fig2}~(c), the more pronounced are the memristive properties of the coupling elements, the higher is the noise intensity being enough for the suppression of travelling waves (values $<R(D)>$ are almost constant for  noise intensities being lower than the threshold value since the initial travelling wave persists). Moreover, at $b\geq1$ the noise-induced suppression occurs very rare and the corresponding curve $<R(D)>$ has barely noticeable or even absent drop. In case $b\geq1$ the dynamics does not principally change when increasing $b$ and the corresponding curves $<R(D)>$ are very close to each other. It is important to note that the noise-induced travelling wave suppression is even not observed at sufficiently large values of $b$ (see the purple curve in Fig.~\ref{fig2}~(c) corresponding to $b=1.5$).

\subsection{L{\'e}vy noise}
%%%%%%%%%%%%%%%%%%%%%%%% FIG 3 %%%%%%%%%%%
\begin{figure}[t!]
\centering
\includegraphics[width=0.48\textwidth]{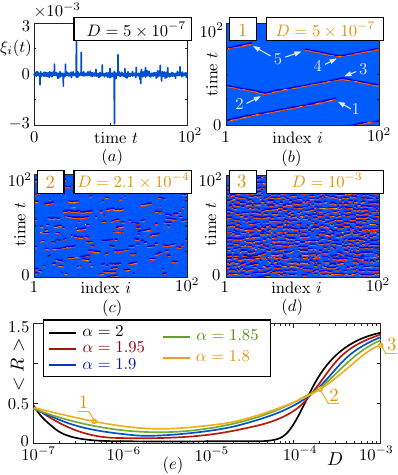}
\caption{(a) Particular realization of L{\'e}vy noise signal $\xi_{i}(t)$ at $\alpha=1.8$, $D=5\times 10^{-7}$. (b)-(d) Transformation of travelling waves when increasing the intensity $D$ of L{\'e}vy noise ($\alpha=1.8$) in the absence of the coupling element memristive properties ($b=0$); (e) The dependencies of the characteristic $<R>$ on the noise intensity $D$ when decreasing the noise scale factor $\alpha$. Points 1-3 in the curve for $\alpha=1.8$ correspond to space-time plots in panels (b)-(d). System parameters are $b=0$, $s=0.7$, $\varepsilon=0.01$.}
\label{fig3}
\end{figure}
%%%%%%%%%%%%%%%%%%%%%%%%%%%%%%%%%%%%%%

In case $\alpha<2$, the noise impact is characterised by the appearance of high-amplitude impulses [Fig.~\ref{fig3}~(a)]. These impulses can both suppress and induce travelling waves. If the coupling does not exhibit the memristive properties ($b=0$) and the noise intensity is low, one can observe both effects on the same spatio-temporal diagram (see points 1-5 in Fig.~\ref{fig3}~(b)). In particular, the initial wave process in Fig.~\ref{fig3}~(b) is suppressed by noise in point 1. However,  two new waves travelling to the left and to the right are induced by noise in point 2 and collide in point 3. Then a new pair of waves is induced in point 4 and destroyed in points 5 due to the action of noise. Increasing the noise intensity leads to the same effects as in case of Gaussian noise: spontaneous spiking activity not associated with travelling waves [Fig.~\ref{fig3}~(c)] and the tendency to synchronization of noise-induced oscillations in the regime of coherence resonance [Fig.~\ref{fig3}~(d)]. As a result of simultaneous suppressing and inducing travelling waves at $\alpha<2$, the sharp drop in the dependency $<R(D)>$ initially observed in the presence of Gaussian noise at $D\in [10^{-7}:3 \times 10^{-7}]$ (see the black curve in Fig.~\ref{fig3}~(e)) becomes smoother and smoother when decreasing the parameter $\alpha$.

%%%%%%%%%%%%%%%%%%%%%%%% FIG 4 %%%%%%%%%%%
\begin{figure}[t!]
\centering
\includegraphics[width=0.48\textwidth]{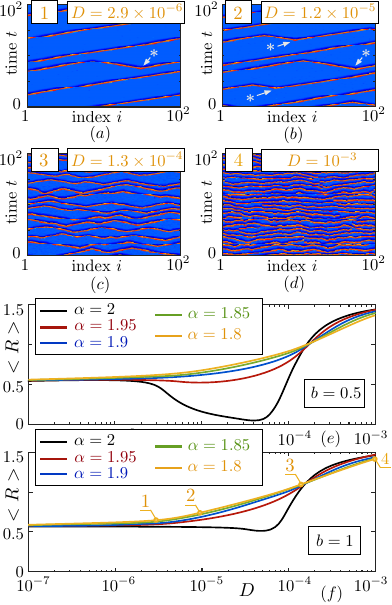}
\caption{(a)-(d) Transformation of travelling waves when increasing the intensity $D$ of L{\'e}vy noise ($\alpha=1.8$) in the presence of the coupling element memristive properties ($b=1$). Panels (e) and (f) illustrate the dependencies $<R(D)>$ when decreasing the noise parameter $\alpha$ and fixing parameter $b$: $b=0.5$ (panel (e)) and $b=1$ (panel (f)). Points 1-4 in panel (f)  correspond to space-time plots in panels (a)-(d). System parameters are $s=0.7$, $\varepsilon=0.01$, $\delta=10^{-2}$.}
\label{fig4}
\end{figure}
%%%%%%%%%%%%%%%%%%%%%%%%%%%%%%%%%%%%%%

Similarly to the case of Gaussian noise, travelling waves are much more robust against the L{\'e}vy noise in the presence of memristive coupling. In particular, high-amplitude noise impulses provide for  appearance of new travelling waves (see points $*$ in Fig.~\ref{fig4}~(a),(b)), whereas spontaneous suppression of waves is not observed even for high noise intensities (see Fig.~\ref{fig4}~(c) where all the noise-induced waves disappear through the collision). Further noise intensity growth transforms the stochastic dynamics into the regime of synchronized spiking activity [Fig.~\ref{fig4}~(d)]. The strengthened travelling wave robustness against the L{\'e}vy noise is reflected in the dependencies $<R(D)>$ obtained when increasing the parameter $b$. In particular, comparative analysis of the curves obtained at $b=0$ [Fig.~\ref{fig3}~(e)], $b=0.5$ [Fig.~\ref{fig4}~(e)], $b=1$ [Fig.~\ref{fig4}~(f)] allows to conclude that the area where the curves $<R(D)>$ have negative slope (where the noise-induced travelling wave destruction is more pronounced than the travelling wave inducing) shrinks and disappears (the dependencies become monotonic) with growth of $b$.

\section{Conclusions}
In the current paper, we have demonstrated a constructive role of memristive coupling in the context of travelling waves observed in an ensemble of deterministic excitable oscillators as well as in the presence of noise. In the deterministic case, the memristive properties of the coupling elements provide for realization of travelling waves at sufficiently lower coupling strength as compared to the classical diffusive interaction of unchangeable intensity. If the travelling waves are exhibited by coupled stochastic oscillators, memristive coupling becomes a factor supporting travelling waves such that the noise-induced destruction of travelling waves is observed at higher noise intensities in comparison with diffusive coupling not exhibiting memristive properties.

The second intriguing effect consists in the action of noise. Depending on the properties of the additive stochastic impact, it can both suppress and induce travelling waves. Noise-induced travelling waves can be easily obtained if the stochastic forcing includes high-amplitude impulses. We used the L{\'e}vy noise model to visualize this fact. 

For sufficiently high noise level, the regime of travelling waves is transformed into the regime of synchronized noise-induced spiking activity being a result of coherence resonance. In our paper, we illustrate this by means of space-time diagrams. However, the revealed evolution into the regime of synchronization requires additional detailed analysis which is an issue for further study. In addition, the presented results can be extended in the following by the consideration of another adaptive coupling models to formulate a generalized conclusion on the impact of synaptic plasticity being an intrinsic property of biological neural networks on effects associated with wave propagation and syncronization in such networks. 

\section*{DATA AVAILABILITY}
The data that support the findings of this study are available from the corresponding author upon reasonable request.

\section*{Acknowledgements}
V.S. and I.K. acknowledge support by the Russian Science Foundation (project No.  23-72-10040).

%\bibliography{bibliography}% Produces the bibliography via BibTeX.

\begin{thebibliography}{62}%
\makeatletter
\providecommand \@ifxundefined [1]{%
 \@ifx{#1\undefined}
}%
\providecommand \@ifnum [1]{%
 \ifnum #1\expandafter \@firstoftwo
 \else \expandafter \@secondoftwo
 \fi
}%
\providecommand \@ifx [1]{%
 \ifx #1\expandafter \@firstoftwo
 \else \expandafter \@secondoftwo
 \fi
}%
\providecommand \natexlab [1]{#1}%
\providecommand \enquote  [1]{``#1''}%
\providecommand \bibnamefont  [1]{#1}%
\providecommand \bibfnamefont [1]{#1}%
\providecommand \citenamefont [1]{#1}%
\providecommand \href@noop [0]{\@secondoftwo}%
\providecommand \href [0]{\begingroup \@sanitize@url \@href}%
\providecommand \@href[1]{\@@startlink{#1}\@@href}%
\providecommand \@@href[1]{\endgroup#1\@@endlink}%
\providecommand \@sanitize@url [0]{\catcode `\\12\catcode `\$12\catcode
  `\&12\catcode `\#12\catcode `\^12\catcode `\_12\catcode `\%12\relax}%
\providecommand \@@startlink[1]{}%
\providecommand \@@endlink[0]{}%
\providecommand \url  [0]{\begingroup\@sanitize@url \@url }%
\providecommand \@url [1]{\endgroup\@href {#1}{\urlprefix }}%
\providecommand \urlprefix  [0]{URL }%
\providecommand \Eprint [0]{\href }%
\providecommand \doibase [0]{http://dx.doi.org/}%
\providecommand \selectlanguage [0]{\@gobble}%
\providecommand \bibinfo  [0]{\@secondoftwo}%
\providecommand \bibfield  [0]{\@secondoftwo}%
\providecommand \translation [1]{[#1]}%
\providecommand \BibitemOpen [0]{}%
\providecommand \bibitemStop [0]{}%
\providecommand \bibitemNoStop [0]{.\EOS\space}%
\providecommand \EOS [0]{\spacefactor3000\relax}%
\providecommand \BibitemShut  [1]{\csname bibitem#1\endcsname}%
\let\auto@bib@innerbib\@empty
%</preamble>
\bibitem [{\citenamefont {Landa}(1996)}]{landa1996}%
  \BibitemOpen
  \bibfield  {author} {\bibinfo {author} {\bibfnamefont {P.}~\bibnamefont
  {Landa}},\ }\href@noop {} {\emph {\bibinfo {title} {Nonlinear Oscillations
  and Waves in Dynamical Systems}}}\ (\bibinfo  {publisher} {Springer},\
  \bibinfo {year} {1996})\BibitemShut {NoStop}%
\bibitem [{\citenamefont {Ghazaryan}, \citenamefont {Lafortune},\ and\
  \citenamefont {Manukian}(2023)}]{ghazaryan2022}%
  \BibitemOpen
  \bibfield  {author} {\bibinfo {author} {\bibfnamefont {A.}~\bibnamefont
  {Ghazaryan}}, \bibinfo {author} {\bibfnamefont {S.}~\bibnamefont
  {Lafortune}}, \ and\ \bibinfo {author} {\bibfnamefont {V.}~\bibnamefont
  {Manukian}},\ }\href@noop {} {\emph {\bibinfo {title} {Introduction to
  Traveling Waves}}}\ (\bibinfo  {publisher} {CRC Press},\ \bibinfo {year}
  {2023})\BibitemShut {NoStop}%
\bibitem [{\citenamefont {Smoller}(1994)}]{smoller1994}%
  \BibitemOpen
  \bibfield  {author} {\bibinfo {author} {\bibfnamefont {J.}~\bibnamefont
  {Smoller}},\ }\href@noop {} {\emph {\bibinfo {title} {Shock Waves and
  Reaction---Diffusion Equations}}}\ (\bibinfo  {publisher} {Springer},\
  \bibinfo {year} {1994})\BibitemShut {NoStop}%
\bibitem [{\citenamefont {Garc{\'\i}a-Ojalvo}\ and\ \citenamefont
  {Sancho}(1999)}]{garcia1999}%
  \BibitemOpen
  \bibfield  {author} {\bibinfo {author} {\bibfnamefont {J.}~\bibnamefont
  {Garc{\'\i}a-Ojalvo}}\ and\ \bibinfo {author} {\bibfnamefont
  {J.}~\bibnamefont {Sancho}},\ }\href@noop {} {\emph {\bibinfo {title} {Noise
  in Spatially Extended Systems}}}\ (\bibinfo  {publisher} {Springer},\
  \bibinfo {year} {1999})\BibitemShut {NoStop}%
\bibitem [{\citenamefont {Shepelev}, \citenamefont {Slepnev},\ and\
  \citenamefont {Vadivasova}(2016)}]{shepelev2016}%
  \BibitemOpen
  \bibfield  {author} {\bibinfo {author} {\bibfnamefont {I.}~\bibnamefont
  {Shepelev}}, \bibinfo {author} {\bibfnamefont {A.}~\bibnamefont {Slepnev}}, \
  and\ \bibinfo {author} {\bibfnamefont {T.}~\bibnamefont {Vadivasova}},\
  }\bibfield  {title} {\enquote {\bibinfo {title} {Different synchronization
  characteristics of distinct types of traveling waves in a model of active
  medium with periodic boundary conditions},}\ }\href@noop {} {\bibfield
  {journal} {\bibinfo  {journal} {Commun. Nonlinear Sci. Numer. Simulat.}\
  }\textbf {\bibinfo {volume} {38}},\ \bibinfo {pages} {206--217} (\bibinfo
  {year} {2016})}\BibitemShut {NoStop}%
\bibitem [{\citenamefont {Korneev}\ \emph
  {et~al.}(2021{\natexlab{a}})\citenamefont {Korneev}, \citenamefont {Semenov},
  \citenamefont {Slepnev},\ and\ \citenamefont {Vadivasova}}]{korneev2021}%
  \BibitemOpen
  \bibfield  {author} {\bibinfo {author} {\bibfnamefont {I.}~\bibnamefont
  {Korneev}}, \bibinfo {author} {\bibfnamefont {V.}~\bibnamefont {Semenov}},
  \bibinfo {author} {\bibfnamefont {A.}~\bibnamefont {Slepnev}}, \ and\
  \bibinfo {author} {\bibfnamefont {T.}~\bibnamefont {Vadivasova}},\ }\bibfield
   {title} {\enquote {\bibinfo {title} {The impact of memristive coupling
  initial states on travelling waves in an ensemble of the Fitzhugh--Nagumo
  oscillators},}\ }\href@noop {} {\bibfield  {journal} {\bibinfo  {journal}
  {Chaos, Solitons and Fractals}\ }\textbf {\bibinfo {volume} {147}},\ \bibinfo
  {pages} {110923} (\bibinfo {year} {2021}{\natexlab{a}})}\BibitemShut
  {NoStop}%
\bibitem [{\citenamefont {Korneev}\ \emph {et~al.}(2022)\citenamefont
  {Korneev}, \citenamefont {Ramazanov}, \citenamefont {Semenov}, \citenamefont
  {Slepnev},\ and\ \citenamefont {Vadivasova}}]{korneev2022}%
  \BibitemOpen
  \bibfield  {author} {\bibinfo {author} {\bibfnamefont {I.}~\bibnamefont
  {Korneev}}, \bibinfo {author} {\bibfnamefont {I.}~\bibnamefont {Ramazanov}},
  \bibinfo {author} {\bibfnamefont {V.}~\bibnamefont {Semenov}}, \bibinfo
  {author} {\bibfnamefont {A.}~\bibnamefont {Slepnev}}, \ and\ \bibinfo
  {author} {\bibfnamefont {T.}~\bibnamefont {Vadivasova}},\ }\bibfield  {title}
  {\enquote {\bibinfo {title} {Synchronization of traveling waves in
  memristively coupled ensembles of Fitzhugh-Nagumo neurons with periodic
  boundary conditions},}\ }\href@noop {} {\bibfield  {journal} {\bibinfo
  {journal} {Frontiers in Physics}\ }\textbf {\bibinfo {volume} {10}},\
  \bibinfo {pages} {886476} (\bibinfo {year} {2022})}\BibitemShut {NoStop}%
\bibitem [{\citenamefont {Semenov}, \citenamefont {Jalan},\ and\ \citenamefont
  {Zakharova}(2023)}]{semenov2023}%
  \BibitemOpen
  \bibfield  {author} {\bibinfo {author} {\bibfnamefont {V.}~\bibnamefont
  {Semenov}}, \bibinfo {author} {\bibfnamefont {S.}~\bibnamefont {Jalan}}, \
  and\ \bibinfo {author} {\bibfnamefont {A.}~\bibnamefont {Zakharova}},\
  }\bibfield  {title} {\enquote {\bibinfo {title} {Multiplexing-based control
  of wavefront propagation: the interplay of inter-layer coupling, asymmetry
  and noise},}\ }\href@noop {} {\bibfield  {journal} {\bibinfo  {journal}
  {Chaos, Solitons and Fractals}\ }\textbf {\bibinfo {volume} {173}},\ \bibinfo
  {pages} {113656} (\bibinfo {year} {2023})}\BibitemShut {NoStop}%
\bibitem [{\citenamefont {Zakharova}\ and\ \citenamefont
  {Semenov}(2023)}]{zakharova2023}%
  \BibitemOpen
  \bibfield  {author} {\bibinfo {author} {\bibfnamefont {A.}~\bibnamefont
  {Zakharova}}\ and\ \bibinfo {author} {\bibfnamefont {V.}~\bibnamefont
  {Semenov}},\ }\bibfield  {title} {\enquote {\bibinfo {title} {Stochastic
  control of spiking activity bump expansion: Monotonic and resonant
  phenomena},}\ }\href@noop {} {\bibfield  {journal} {\bibinfo  {journal}
  {Chaos}\ }\textbf {\bibinfo {volume} {33}},\ \bibinfo {pages} {081101}
  (\bibinfo {year} {2023})}\BibitemShut {NoStop}%
\bibitem [{\citenamefont {Rybalova}, \citenamefont {Muni},\ and\ \citenamefont
  {Strelkova}(2023)}]{rybalova2023}%
  \BibitemOpen
  \bibfield  {author} {\bibinfo {author} {\bibfnamefont {E.}~\bibnamefont
  {Rybalova}}, \bibinfo {author} {\bibfnamefont {S.}~\bibnamefont {Muni}}, \
  and\ \bibinfo {author} {\bibfnamefont {G.}~\bibnamefont {Strelkova}},\
  }\bibfield  {title} {\enquote {\bibinfo {title} {Transition from
  chimera/solitary states to traveling waves},}\ }\href@noop {} {\bibfield
  {journal} {\bibinfo  {journal} {Chaos}\ }\textbf {\bibinfo {volume} {33}},\
  \bibinfo {pages} {033104} (\bibinfo {year} {2023})}\BibitemShut {NoStop}%
\bibitem [{\citenamefont {Giacomelli}\ \emph {et~al.}(2012)\citenamefont
  {Giacomelli}, \citenamefont {Marino}, \citenamefont {Zaks},\ and\
  \citenamefont {Yanchuk}}]{giacomelli2012}%
  \BibitemOpen
  \bibfield  {author} {\bibinfo {author} {\bibfnamefont {G.}~\bibnamefont
  {Giacomelli}}, \bibinfo {author} {\bibfnamefont {F.}~\bibnamefont {Marino}},
  \bibinfo {author} {\bibfnamefont {M.~A.}\ \bibnamefont {Zaks}}, \ and\
  \bibinfo {author} {\bibfnamefont {S.}~\bibnamefont {Yanchuk}},\ }\bibfield
  {title} {\enquote {\bibinfo {title} {Coarsening in a bistable system with
  long-delayed feedback},}\ }\href@noop {} {\bibfield  {journal} {\bibinfo
  {journal} {Europhys. Lett.}\ }\textbf {\bibinfo {volume} {99}},\ \bibinfo
  {pages} {58005} (\bibinfo {year} {2012})}\BibitemShut {NoStop}%
\bibitem [{\citenamefont {Klinshov}\ \emph {et~al.}(2017)\citenamefont
  {Klinshov}, \citenamefont {Shchapin}, \citenamefont {Yanchuk}, \citenamefont
  {Wolfrum}, \citenamefont {D'Huys},\ and\ \citenamefont
  {Nekorkin}}]{klinshov2017}%
  \BibitemOpen
  \bibfield  {author} {\bibinfo {author} {\bibfnamefont {V.}~\bibnamefont
  {Klinshov}}, \bibinfo {author} {\bibfnamefont {D.}~\bibnamefont {Shchapin}},
  \bibinfo {author} {\bibfnamefont {S.}~\bibnamefont {Yanchuk}}, \bibinfo
  {author} {\bibfnamefont {M.}~\bibnamefont {Wolfrum}}, \bibinfo {author}
  {\bibfnamefont {O.}~\bibnamefont {D'Huys}}, \ and\ \bibinfo {author}
  {\bibfnamefont {V.}~\bibnamefont {Nekorkin}},\ }\bibfield  {title} {\enquote
  {\bibinfo {title} {Embedding the dynamics of a single delay system into a
  feed-forward ring},}\ }\href@noop {} {\bibfield  {journal} {\bibinfo
  {journal} {Phys. Rev. E}\ }\textbf {\bibinfo {volume} {96}},\ \bibinfo
  {pages} {042217} (\bibinfo {year} {2017})}\BibitemShut {NoStop}%
\bibitem [{\citenamefont {Semenov}\ and\ \citenamefont
  {Maistrenko}(2018)}]{semenov2018}%
  \BibitemOpen
  \bibfield  {author} {\bibinfo {author} {\bibfnamefont {V.}~\bibnamefont
  {Semenov}}\ and\ \bibinfo {author} {\bibfnamefont {Y.}~\bibnamefont
  {Maistrenko}},\ }\bibfield  {title} {\enquote {\bibinfo {title} {Dissipative
  solitons for bistable delayed-feedback systems},}\ }\href@noop {} {\bibfield
  {journal} {\bibinfo  {journal} {Chaos}\ }\textbf {\bibinfo {volume} {28}},\
  \bibinfo {pages} {101103} (\bibinfo {year} {2018})}\BibitemShut {NoStop}%
\bibitem [{\citenamefont {Zakharova}\ and\ \citenamefont
  {Semenov}(2024)}]{zakharova2024}%
  \BibitemOpen
  \bibfield  {author} {\bibinfo {author} {\bibfnamefont {A.}~\bibnamefont
  {Zakharova}}\ and\ \bibinfo {author} {\bibfnamefont {V.}~\bibnamefont
  {Semenov}},\ }\bibfield  {title} {\enquote {\bibinfo {title}
  {Delayed-feedback oscillators replicate the dynamics of multiplex networks:
  wavefront propagation and stochastic resonance},}\ }\href@noop {} {\bibfield
  {journal} {\bibinfo  {journal} {arXiv:}\ }\textbf {\bibinfo {volume}
  {2402.16551}} (\bibinfo {year} {2024})}\BibitemShut {NoStop}%
\bibitem [{\citenamefont {Loarte}\ \emph {et~al.}(1999)\citenamefont {Loarte},
  \citenamefont {Monk}, \citenamefont {Kukushkin}, \citenamefont {Righi},
  \citenamefont {Campbell}, \citenamefont {Conway},\ and\ \citenamefont
  {Maggi}}]{loarte1999}%
  \BibitemOpen
  \bibfield  {author} {\bibinfo {author} {\bibfnamefont {A.}~\bibnamefont
  {Loarte}}, \bibinfo {author} {\bibfnamefont {R.~D.}\ \bibnamefont {Monk}},
  \bibinfo {author} {\bibfnamefont {A.~S.}\ \bibnamefont {Kukushkin}}, \bibinfo
  {author} {\bibfnamefont {E.}~\bibnamefont {Righi}}, \bibinfo {author}
  {\bibfnamefont {D.~J.}\ \bibnamefont {Campbell}}, \bibinfo {author}
  {\bibfnamefont {G.~D.}\ \bibnamefont {Conway}}, \ and\ \bibinfo {author}
  {\bibfnamefont {C.~F.}\ \bibnamefont {Maggi}},\ }\bibfield  {title} {\enquote
  {\bibinfo {title} {Self-sustained divertor plasma oscillations in the jet
  tokamak},}\ }\href@noop {} {\bibfield  {journal} {\bibinfo  {journal} {Phys.
  Rev. Lett.}\ }\textbf {\bibinfo {volume} {83}},\ \bibinfo {pages}
  {3657--3660} (\bibinfo {year} {1999})}\BibitemShut {NoStop}%
\bibitem [{\citenamefont {Schwabe}\ \emph {et~al.}(2007)\citenamefont
  {Schwabe}, \citenamefont {Rubin-Zuzic}, \citenamefont {Zhdanov},
  \citenamefont {Thomas},\ and\ \citenamefont {Morfill}}]{schwabe2007}%
  \BibitemOpen
  \bibfield  {author} {\bibinfo {author} {\bibfnamefont {M.}~\bibnamefont
  {Schwabe}}, \bibinfo {author} {\bibfnamefont {M.}~\bibnamefont
  {Rubin-Zuzic}}, \bibinfo {author} {\bibfnamefont {S.}~\bibnamefont
  {Zhdanov}}, \bibinfo {author} {\bibfnamefont {H.~M.}\ \bibnamefont {Thomas}},
  \ and\ \bibinfo {author} {\bibfnamefont {G.~E.}\ \bibnamefont {Morfill}},\
  }\bibfield  {title} {\enquote {\bibinfo {title} {Highly resolved self-excited
  density waves in a complex plasma},}\ }\href@noop {} {\bibfield  {journal}
  {\bibinfo  {journal} {Phys. Rev. Lett.}\ }\textbf {\bibinfo {volume} {99}},\
  \bibinfo {pages} {095002} (\bibinfo {year} {2007})}\BibitemShut {NoStop}%
\bibitem [{\citenamefont {Ebeling}(1993)}]{ebeling1993}%
  \BibitemOpen
  \bibfield  {author} {\bibinfo {author} {\bibfnamefont {K.}~\bibnamefont
  {Ebeling}},\ }\href@noop {} {\emph {\bibinfo {title} {Integrated
  Optoelectronics}}}\ (\bibinfo  {publisher} {Springer},\ \bibinfo {year}
  {1993})\BibitemShut {NoStop}%
\bibitem [{\citenamefont {Akhmediev}\ and\ \citenamefont
  {Ankiewicz}(1997)}]{akhmediev1997}%
  \BibitemOpen
  \bibfield  {author} {\bibinfo {author} {\bibfnamefont {N.}~\bibnamefont
  {Akhmediev}}\ and\ \bibinfo {author} {\bibfnamefont {A.}~\bibnamefont
  {Ankiewicz}},\ }\href@noop {} {\emph {\bibinfo {title} {Solitons}}}\
  (\bibinfo  {publisher} {Springer},\ \bibinfo {year} {1997})\BibitemShut
  {NoStop}%
\bibitem [{\citenamefont {Semenov}\ \emph {et~al.}(2023)\citenamefont
  {Semenov}, \citenamefont {Porte}, \citenamefont {Larger},\ and\ \citenamefont
  {Brunner}}]{semenov2023-2}%
  \BibitemOpen
  \bibfield  {author} {\bibinfo {author} {\bibfnamefont {V.}~\bibnamefont
  {Semenov}}, \bibinfo {author} {\bibfnamefont {X.}~\bibnamefont {Porte}},
  \bibinfo {author} {\bibfnamefont {L.}~\bibnamefont {Larger}}, \ and\ \bibinfo
  {author} {\bibfnamefont {D.}~\bibnamefont {Brunner}},\ }\bibfield  {title}
  {\enquote {\bibinfo {title} {Deterministic and stochastic coarsening control
  in optically-addressed spatial light modulators subject to optical
  feedback},}\ }\href@noop {} {\bibfield  {journal} {\bibinfo  {journal}
  {Physical Review B}\ }\textbf {\bibinfo {volume} {108}},\ \bibinfo {pages}
  {024307} (\bibinfo {year} {2023})}\BibitemShut {NoStop}%
\bibitem [{\citenamefont {Liu}(2016)}]{liu2016}%
  \BibitemOpen
  \bibfield  {author} {\bibinfo {author} {\bibfnamefont {J.-M.}\ \bibnamefont
  {Liu}},\ }\href@noop {} {\emph {\bibinfo {title} {Principles of photonics}}}\
  (\bibinfo  {publisher} {Cambridge University Press},\ \bibinfo {year}
  {2016})\BibitemShut {NoStop}%
\bibitem [{\citenamefont {Henry}\ \emph {et~al.}(2019)\citenamefont {Henry},
  \citenamefont {Kalimeris}, \citenamefont {P{\u a}r{\u a}u}, \citenamefont
  {Vanden-Broeck},\ and\ \citenamefont {Wahl{\'e}n}}]{henry2019}%
  \BibitemOpen
  \bibinfo {editor} {\bibfnamefont {D.}~\bibnamefont {Henry}}, \bibinfo
  {editor} {\bibfnamefont {K.}~\bibnamefont {Kalimeris}}, \bibinfo {editor}
  {\bibfnamefont {E.}~\bibnamefont {P{\u a}r{\u a}u}}, \bibinfo {editor}
  {\bibfnamefont {J.-M.}\ \bibnamefont {Vanden-Broeck}}, \ and\ \bibinfo
  {editor} {\bibfnamefont {E.}~\bibnamefont {Wahl{\'e}n}},\ eds.,\ \href@noop
  {} {\emph {\bibinfo {title} {Nonlinear Water Waves}}}\ (\bibinfo  {publisher}
  {Birkh{\"a}user},\ \bibinfo {year} {2019})\BibitemShut {NoStop}%
\bibitem [{\citenamefont {Debnath}(1994)}]{debnath1994}%
  \BibitemOpen
  \bibfield  {author} {\bibinfo {author} {\bibfnamefont {L.}~\bibnamefont
  {Debnath}},\ }\href@noop {} {\emph {\bibinfo {title} {Nonlinear Water
  Waves}}}\ (\bibinfo  {publisher} {Academic Press, Boston},\ \bibinfo {year}
  {1994})\BibitemShut {NoStop}%
\bibitem [{\citenamefont {Johnson}(1997)}]{johnson1997}%
  \BibitemOpen
  \bibfield  {author} {\bibinfo {author} {\bibfnamefont {R.}~\bibnamefont
  {Johnson}},\ }\href@noop {} {\emph {\bibinfo {title} {A Modern Introduction
  to the Mathematical Theory of Water Waves}}}\ (\bibinfo  {publisher}
  {Cambridge University Press},\ \bibinfo {year} {1997})\BibitemShut {NoStop}%
\bibitem [{\citenamefont {Dias}\ and\ \citenamefont {Kharif}(1999)}]{dias1999}%
  \BibitemOpen
  \bibfield  {author} {\bibinfo {author} {\bibfnamefont {F.}~\bibnamefont
  {Dias}}\ and\ \bibinfo {author} {\bibfnamefont {C.}~\bibnamefont {Kharif}},\
  }\bibfield  {title} {\enquote {\bibinfo {title} {Nonlinear gravity and
  capillary-gravity waves},}\ }\href@noop {} {\bibfield  {journal} {\bibinfo
  {journal} {Annual Review of Fluid Mechanics}\ }\textbf {\bibinfo {volume}
  {31}},\ \bibinfo {pages} {301--346} (\bibinfo {year} {1999})}\BibitemShut
  {NoStop}%
\bibitem [{\citenamefont {Kuramoto}(1984)}]{kuramoto1984}%
  \BibitemOpen
  \bibfield  {author} {\bibinfo {author} {\bibfnamefont {Y.}~\bibnamefont
  {Kuramoto}},\ }\href@noop {} {\emph {\bibinfo {title} {Chemical Oscillations,
  Waves, and Turbulence}}}\ (\bibinfo  {publisher} {Springer},\ \bibinfo {year}
  {1984})\BibitemShut {NoStop}%
\bibitem [{\citenamefont {Kapral}\ and\ \citenamefont
  {Showalter}(1995)}]{kapral1995}%
  \BibitemOpen
  \bibinfo {editor} {\bibfnamefont {R.}~\bibnamefont {Kapral}}\ and\ \bibinfo
  {editor} {\bibfnamefont {K.}~\bibnamefont {Showalter}},\ eds.,\ \href@noop {}
  {\emph {\bibinfo {title} {Chemical Waves and Patterns}}}\ (\bibinfo
  {publisher} {Springer},\ \bibinfo {year} {1995})\BibitemShut {NoStop}%
\bibitem [{\citenamefont {Epstein}\ and\ \citenamefont
  {Pojman}(1998)}]{epstein1998}%
  \BibitemOpen
  \bibfield  {author} {\bibinfo {author} {\bibfnamefont {I.}~\bibnamefont
  {Epstein}}\ and\ \bibinfo {author} {\bibfnamefont {J.}~\bibnamefont
  {Pojman}},\ }\href@noop {} {\emph {\bibinfo {title} {An Introduction to
  Nonlinear Chemical Dynamics: Oscillations, Waves, Patterns, and Chaos}}}\
  (\bibinfo  {publisher} {Oxford University Press},\ \bibinfo {year}
  {1998})\BibitemShut {NoStop}%
\bibitem [{\citenamefont {Hugles}(1995)}]{hugles1995}%
  \BibitemOpen
  \bibfield  {author} {\bibinfo {author} {\bibfnamefont {J.}~\bibnamefont
  {Hugles}},\ }\bibfield  {title} {\enquote {\bibinfo {title} {The phenomenon
  of travelling waves: A review},}\ }\href@noop {} {\bibfield  {journal}
  {\bibinfo  {journal} {Clinical Electroencephalography}\ }\textbf {\bibinfo
  {volume} {26}},\ \bibinfo {pages} {1--6} (\bibinfo {year}
  {1995})}\BibitemShut {NoStop}%
\bibitem [{\citenamefont {Muller}\ \emph {et~al.}(2018)\citenamefont {Muller},
  \citenamefont {Chavane}, \citenamefont {Reynolds},\ and\ \citenamefont
  {Sejnowski}}]{muller2018}%
  \BibitemOpen
  \bibfield  {author} {\bibinfo {author} {\bibfnamefont {L.}~\bibnamefont
  {Muller}}, \bibinfo {author} {\bibfnamefont {F.}~\bibnamefont {Chavane}},
  \bibinfo {author} {\bibfnamefont {J.}~\bibnamefont {Reynolds}}, \ and\
  \bibinfo {author} {\bibfnamefont {T.}~\bibnamefont {Sejnowski}},\ }\bibfield
  {title} {\enquote {\bibinfo {title} {Cortical travelling waves: mechanisms
  and computational principles},}\ }\href@noop {} {\bibfield  {journal}
  {\bibinfo  {journal} {Nature Reviews Neuroscience}\ }\textbf {\bibinfo
  {volume} {19}},\ \bibinfo {pages} {255--268} (\bibinfo {year}
  {2018})}\BibitemShut {NoStop}%
\bibitem [{\citenamefont {Mohan}\ \emph {et~al.}(2024)\citenamefont {Mohan},
  \citenamefont {Zhang}, \citenamefont {Ermentrout},\ and\ \citenamefont
  {Jacobs}}]{mohan2024}%
  \BibitemOpen
  \bibfield  {author} {\bibinfo {author} {\bibfnamefont {U.}~\bibnamefont
  {Mohan}}, \bibinfo {author} {\bibfnamefont {H.}~\bibnamefont {Zhang}},
  \bibinfo {author} {\bibfnamefont {B.}~\bibnamefont {Ermentrout}}, \ and\
  \bibinfo {author} {\bibfnamefont {J.}~\bibnamefont {Jacobs}},\ }\bibfield
  {title} {\enquote {\bibinfo {title} {The direction of theta and alpha
  travelling waves modulates human memory processing},}\ }\href
  {https://doi.org/10.1038/s41562-024-01838-3} {\bibfield  {journal} {\bibinfo
  {journal} {Nature Human Behaviour}\ } (\bibinfo {year} {2024})}\BibitemShut
  {NoStop}%
\bibitem [{\citenamefont {Sherratt}\ and\ \citenamefont
  {Smith}(2008)}]{sherratt2008}%
  \BibitemOpen
  \bibfield  {author} {\bibinfo {author} {\bibfnamefont {J.}~\bibnamefont
  {Sherratt}}\ and\ \bibinfo {author} {\bibfnamefont {M.}~\bibnamefont
  {Smith}},\ }\bibfield  {title} {\enquote {\bibinfo {title} {Periodic
  travelling waves in cyclic populations: field studies and reaction--diffusion
  models},}\ }\href@noop {} {\bibfield  {journal} {\bibinfo  {journal} {J. R.
  Soc. Interface}\ }\textbf {\bibinfo {volume} {5}},\ \bibinfo {pages}
  {483--505} (\bibinfo {year} {2008})}\BibitemShut {NoStop}%
\bibitem [{\citenamefont {Malchow}, \citenamefont {Petrovskii},\ and\
  \citenamefont {Venturino}(2008)}]{malchow2008}%
  \BibitemOpen
  \bibfield  {author} {\bibinfo {author} {\bibfnamefont {H.}~\bibnamefont
  {Malchow}}, \bibinfo {author} {\bibfnamefont {S.}~\bibnamefont {Petrovskii}},
  \ and\ \bibinfo {author} {\bibfnamefont {E.}~\bibnamefont {Venturino}},\
  }\href@noop {} {\emph {\bibinfo {title} {Spatiotemporal Patterns in Ecology
  and Epidemiology}}}\ (\bibinfo  {publisher} {CRC Press},\ \bibinfo {year}
  {2008})\BibitemShut {NoStop}%
\bibitem [{\citenamefont {Brauer}\ and\ \citenamefont
  {Castillo-Chavez}(2012)}]{brauer2012}%
  \BibitemOpen
  \bibfield  {author} {\bibinfo {author} {\bibfnamefont {F.}~\bibnamefont
  {Brauer}}\ and\ \bibinfo {author} {\bibfnamefont {C.}~\bibnamefont
  {Castillo-Chavez}},\ }\href@noop {} {\emph {\bibinfo {title} {Mathematical
  Models in Population Biology and Epidemiology}}}\ (\bibinfo  {publisher}
  {Springer},\ \bibinfo {year} {2012})\BibitemShut {NoStop}%
\bibitem [{\citenamefont {Kozma}, \citenamefont {Pino},\ and\ \citenamefont
  {Pazienza}(2012)}]{kozma2012}%
  \BibitemOpen
  \bibinfo {editor} {\bibfnamefont {R.}~\bibnamefont {Kozma}}, \bibinfo
  {editor} {\bibfnamefont {R.}~\bibnamefont {Pino}}, \ and\ \bibinfo {editor}
  {\bibfnamefont {G.}~\bibnamefont {Pazienza}},\ eds.,\ \href@noop {} {\emph
  {\bibinfo {title} {Advances in Neuromorphic Memristor Science and
  Applications}}}\ (\bibinfo  {publisher} {Springer},\ \bibinfo {year}
  {2012})\BibitemShut {NoStop}%
\bibitem [{\citenamefont {Chua}, \citenamefont {Tetzlaff},\ and\ \citenamefont
  {Slavova}(2022)}]{chua2022}%
  \BibitemOpen
  \bibinfo {editor} {\bibfnamefont {L.}~\bibnamefont {Chua}}, \bibinfo {editor}
  {\bibfnamefont {R.}~\bibnamefont {Tetzlaff}}, \ and\ \bibinfo {editor}
  {\bibfnamefont {A.}~\bibnamefont {Slavova}},\ eds.,\ \href@noop {} {\emph
  {\bibinfo {title} {Memristor Computing Systems}}}\ (\bibinfo  {publisher}
  {Springer},\ \bibinfo {year} {2022})\BibitemShut {NoStop}%
\bibitem [{\citenamefont {Korneev}, \citenamefont {Vadivasova},\ and\
  \citenamefont {Semenov}(2017)}]{korneev2017}%
  \BibitemOpen
  \bibfield  {author} {\bibinfo {author} {\bibfnamefont {I.}~\bibnamefont
  {Korneev}}, \bibinfo {author} {\bibfnamefont {T.}~\bibnamefont {Vadivasova}},
  \ and\ \bibinfo {author} {\bibfnamefont {V.}~\bibnamefont {Semenov}},\
  }\bibfield  {title} {\enquote {\bibinfo {title} {Hard and soft excitation of
  oscillations in memristor-based oscillators with a line of equilibria},}\
  }\href@noop {} {\bibfield  {journal} {\bibinfo  {journal} {Nonlinear
  Dynamics}\ }\textbf {\bibinfo {volume} {89}},\ \bibinfo {pages} {2829--2843}
  (\bibinfo {year} {2017})}\BibitemShut {NoStop}%
\bibitem [{\citenamefont {Korneev}\ and\ \citenamefont
  {Semenov}(2017)}]{korneev2017-2}%
  \BibitemOpen
  \bibfield  {author} {\bibinfo {author} {\bibfnamefont {I.}~\bibnamefont
  {Korneev}}\ and\ \bibinfo {author} {\bibfnamefont {V.}~\bibnamefont
  {Semenov}},\ }\bibfield  {title} {\enquote {\bibinfo {title} {Andronov-Hopf
  bifurcation with and without parameter in a cubic memristor oscillator with a
  line of equilibria},}\ }\href@noop {} {\bibfield  {journal} {\bibinfo
  {journal} {Chaos}\ }\textbf {\bibinfo {volume} {27}},\ \bibinfo {pages}
  {081104} (\bibinfo {year} {2017})}\BibitemShut {NoStop}%
\bibitem [{\citenamefont {Korneev}\ \emph
  {et~al.}(2021{\natexlab{b}})\citenamefont {Korneev}, \citenamefont {Slepnev},
  \citenamefont {Vadivasova},\ and\ \citenamefont {Semenov}}]{korneev2021-2}%
  \BibitemOpen
  \bibfield  {author} {\bibinfo {author} {\bibfnamefont {I.}~\bibnamefont
  {Korneev}}, \bibinfo {author} {\bibfnamefont {A.}~\bibnamefont {Slepnev}},
  \bibinfo {author} {\bibfnamefont {T.}~\bibnamefont {Vadivasova}}, \ and\
  \bibinfo {author} {\bibfnamefont {V.}~\bibnamefont {Semenov}},\ }\bibfield
  {title} {\enquote {\bibinfo {title} {Subcritical Andronov-Hopf scenario for
  systems with a line of equilibria},}\ }\href@noop {} {\bibfield  {journal}
  {\bibinfo  {journal} {Chaos}\ }\textbf {\bibinfo {volume} {31}},\ \bibinfo
  {pages} {073102} (\bibinfo {year} {2021}{\natexlab{b}})}\BibitemShut
  {NoStop}%
\bibitem [{\citenamefont {Korneev}\ \emph
  {et~al.}(2023{\natexlab{a}})\citenamefont {Korneev}, \citenamefont {Slepnev},
  \citenamefont {Zakharova}, \citenamefont {Vadivasova},\ and\ \citenamefont
  {Semenov}}]{korneev2023}%
  \BibitemOpen
  \bibfield  {author} {\bibinfo {author} {\bibfnamefont {I.}~\bibnamefont
  {Korneev}}, \bibinfo {author} {\bibfnamefont {A.}~\bibnamefont {Slepnev}},
  \bibinfo {author} {\bibfnamefont {A.}~\bibnamefont {Zakharova}}, \bibinfo
  {author} {\bibfnamefont {T.}~\bibnamefont {Vadivasova}}, \ and\ \bibinfo
  {author} {\bibfnamefont {V.}~\bibnamefont {Semenov}},\ }\bibfield  {title}
  {\enquote {\bibinfo {title} {Generalized model for steady-state bifurcations
  without parameters in memristor-based oscillators with lines of
  equilibria},}\ }\href@noop {} {\bibfield  {journal} {\bibinfo  {journal}
  {Nonlinear Dynamics}\ }\textbf {\bibinfo {volume} {111}},\ \bibinfo {pages}
  {1235--1243} (\bibinfo {year} {2023}{\natexlab{a}})}\BibitemShut {NoStop}%
\bibitem [{\citenamefont {Korneev}\ \emph
  {et~al.}(2023{\natexlab{b}})\citenamefont {Korneev}, \citenamefont
  {Ramazanov}, \citenamefont {Slepnev}, \citenamefont {Vadivasova},\ and\
  \citenamefont {Semenov}}]{korneev2023-2}%
  \BibitemOpen
  \bibfield  {author} {\bibinfo {author} {\bibfnamefont {I.}~\bibnamefont
  {Korneev}}, \bibinfo {author} {\bibfnamefont {I.}~\bibnamefont {Ramazanov}},
  \bibinfo {author} {\bibfnamefont {A.}~\bibnamefont {Slepnev}}, \bibinfo
  {author} {\bibfnamefont {T.}~\bibnamefont {Vadivasova}}, \ and\ \bibinfo
  {author} {\bibfnamefont {V.}~\bibnamefont {Semenov}},\ }\bibfield  {title}
  {\enquote {\bibinfo {title} {Feigenbaum scenario without parameters},}\
  }\href@noop {} {\bibfield  {journal} {\bibinfo  {journal} {Chaos}\ }\textbf
  {\bibinfo {volume} {33}},\ \bibinfo {pages} {091102} (\bibinfo {year}
  {2023}{\natexlab{b}})}\BibitemShut {NoStop}%
\bibitem [{\citenamefont {Buscarino}\ \emph {et~al.}(2016)\citenamefont
  {Buscarino}, \citenamefont {Corradino}, \citenamefont {Fortuna},
  \citenamefont {Frasca},\ and\ \citenamefont {Chua}}]{buscarino2016}%
  \BibitemOpen
  \bibfield  {author} {\bibinfo {author} {\bibfnamefont {A.}~\bibnamefont
  {Buscarino}}, \bibinfo {author} {\bibfnamefont {C.}~\bibnamefont
  {Corradino}}, \bibinfo {author} {\bibfnamefont {L.}~\bibnamefont {Fortuna}},
  \bibinfo {author} {\bibfnamefont {M.}~\bibnamefont {Frasca}}, \ and\ \bibinfo
  {author} {\bibfnamefont {L.}~\bibnamefont {Chua}},\ }\bibfield  {title}
  {\enquote {\bibinfo {title} {Turing patterns in memristive cellular nonlinear
  networks},}\ }\href@noop {} {\bibfield  {journal} {\bibinfo  {journal} {IEEE
  Trans. on Circuits and Systems}\ }\textbf {\bibinfo {volume} {PP}},\ \bibinfo
  {pages} {1--9} (\bibinfo {year} {2016})}\BibitemShut {NoStop}%
\bibitem [{\citenamefont {Pham}\ \emph {et~al.}(2012)\citenamefont {Pham},
  \citenamefont {Buscarino}, \citenamefont {Fortuna},\ and\ \citenamefont
  {Frasca}}]{pham2012}%
  \BibitemOpen
  \bibfield  {author} {\bibinfo {author} {\bibfnamefont {V.-T.}\ \bibnamefont
  {Pham}}, \bibinfo {author} {\bibfnamefont {A.}~\bibnamefont {Buscarino}},
  \bibinfo {author} {\bibfnamefont {L.}~\bibnamefont {Fortuna}}, \ and\
  \bibinfo {author} {\bibfnamefont {M.}~\bibnamefont {Frasca}},\ }\bibfield
  {title} {\enquote {\bibinfo {title} {Autowaves in memristive cellular neural
  networks},}\ }\href@noop {} {\bibfield  {journal} {\bibinfo  {journal}
  {International Journal of Bifurcation and Chaos}\ }\textbf {\bibinfo {volume}
  {22}},\ \bibinfo {pages} {1230027} (\bibinfo {year} {2012})}\BibitemShut
  {NoStop}%
\bibitem [{\citenamefont {Korneev}, \citenamefont {Semenov},\ and\
  \citenamefont {Vadivasova}(2020)}]{korneev2020}%
  \BibitemOpen
  \bibfield  {author} {\bibinfo {author} {\bibfnamefont {I.}~\bibnamefont
  {Korneev}}, \bibinfo {author} {\bibfnamefont {V.}~\bibnamefont {Semenov}}, \
  and\ \bibinfo {author} {\bibfnamefont {T.}~\bibnamefont {Vadivasova}},\
  }\bibfield  {title} {\enquote {\bibinfo {title} {Synchronization of periodic
  self-oscillators interacting via memristor-based coupling},}\ }\href@noop {}
  {\bibfield  {journal} {\bibinfo  {journal} {International Journal of
  Bifurcation and Chaos}\ }\textbf {\bibinfo {volume} {30}},\ \bibinfo {pages}
  {2050096} (\bibinfo {year} {2020})}\BibitemShut {NoStop}%
\bibitem [{\citenamefont {Korneev}\ \emph
  {et~al.}(2021{\natexlab{c}})\citenamefont {Korneev}, \citenamefont {Semenov},
  \citenamefont {Slepnev},\ and\ \citenamefont {Vadivasova}}]{korneev2021-3}%
  \BibitemOpen
  \bibfield  {author} {\bibinfo {author} {\bibfnamefont {I.}~\bibnamefont
  {Korneev}}, \bibinfo {author} {\bibfnamefont {V.}~\bibnamefont {Semenov}},
  \bibinfo {author} {\bibfnamefont {A.}~\bibnamefont {Slepnev}}, \ and\
  \bibinfo {author} {\bibfnamefont {T.}~\bibnamefont {Vadivasova}},\ }\bibfield
   {title} {\enquote {\bibinfo {title} {Complete synchronization of chaos in
  systems with nonlinear inertial coupling},}\ }\href@noop {} {\bibfield
  {journal} {\bibinfo  {journal} {Chaos, Solitons and Fractals}\ }\textbf
  {\bibinfo {volume} {142}},\ \bibinfo {pages} {110459} (\bibinfo {year}
  {2021}{\natexlab{c}})}\BibitemShut {NoStop}%
\bibitem [{\citenamefont {Jo}\ \emph {et~al.}(2010)\citenamefont {Jo},
  \citenamefont {Chang}, \citenamefont {Ebong}, \citenamefont {Bhadviya},
  \citenamefont {Mazumder},\ and\ \citenamefont {Lu}}]{jo2010}%
  \BibitemOpen
  \bibfield  {author} {\bibinfo {author} {\bibfnamefont {S.}~\bibnamefont
  {Jo}}, \bibinfo {author} {\bibfnamefont {T.}~\bibnamefont {Chang}}, \bibinfo
  {author} {\bibfnamefont {I.}~\bibnamefont {Ebong}}, \bibinfo {author}
  {\bibfnamefont {B.}~\bibnamefont {Bhadviya}}, \bibinfo {author}
  {\bibfnamefont {P.}~\bibnamefont {Mazumder}}, \ and\ \bibinfo {author}
  {\bibfnamefont {W.}~\bibnamefont {Lu}},\ }\bibfield  {title} {\enquote
  {\bibinfo {title} {Nanoscale memristor device as synapse in neuromorphic
  systems},}\ }\href@noop {} {\bibfield  {journal} {\bibinfo  {journal} {Nano
  Lett.}\ }\textbf {\bibinfo {volume} {10}},\ \bibinfo {pages} {1297--1301}
  (\bibinfo {year} {2010})}\BibitemShut {NoStop}%
\bibitem [{\citenamefont {Li}\ \emph {et~al.}(2013)\citenamefont {Li},
  \citenamefont {Zhong}, \citenamefont {Xu}, \citenamefont {Zhang},
  \citenamefont {Xu},\ and\ \citenamefont {Miao}}]{li2013}%
  \BibitemOpen
  \bibfield  {author} {\bibinfo {author} {\bibfnamefont {Y.}~\bibnamefont
  {Li}}, \bibinfo {author} {\bibfnamefont {Y.}~\bibnamefont {Zhong}}, \bibinfo
  {author} {\bibfnamefont {L.}~\bibnamefont {Xu}}, \bibinfo {author}
  {\bibfnamefont {J.}~\bibnamefont {Zhang}}, \bibinfo {author} {\bibfnamefont
  {H.}~\bibnamefont {Xu}, \bibfnamefont {X.~Sun}}, \ and\ \bibinfo {author}
  {\bibfnamefont {X.}~\bibnamefont {Miao}},\ }\bibfield  {title} {\enquote
  {\bibinfo {title} {Ultrafast synaptic events in a chalcogenide memristor},}\
  }\href@noop {} {\bibfield  {journal} {\bibinfo  {journal} {Scientific
  Reports}\ }\textbf {\bibinfo {volume} {3}},\ \bibinfo {pages} {1619}
  (\bibinfo {year} {2013})}\BibitemShut {NoStop}%
\bibitem [{\citenamefont {Serb}\ \emph {et~al.}(2016)\citenamefont {Serb},
  \citenamefont {Bill}, \citenamefont {Khiat}, \citenamefont {Berdan},
  \citenamefont {Legenstein},\ and\ \citenamefont {Prodromakis}}]{serb2016}%
  \BibitemOpen
  \bibfield  {author} {\bibinfo {author} {\bibfnamefont {A.}~\bibnamefont
  {Serb}}, \bibinfo {author} {\bibfnamefont {J.}~\bibnamefont {Bill}}, \bibinfo
  {author} {\bibfnamefont {A.}~\bibnamefont {Khiat}}, \bibinfo {author}
  {\bibfnamefont {R.}~\bibnamefont {Berdan}}, \bibinfo {author} {\bibfnamefont
  {R.}~\bibnamefont {Legenstein}}, \ and\ \bibinfo {author} {\bibfnamefont
  {T.}~\bibnamefont {Prodromakis}},\ }\bibfield  {title} {\enquote {\bibinfo
  {title} {Unsupervised learning in probabilistic neural networks with
  multi-state metal-oxide memristive synapses},}\ }\href@noop {} {\bibfield
  {journal} {\bibinfo  {journal} {Nature Communications}\ }\textbf {\bibinfo
  {volume} {7}},\ \bibinfo {pages} {12611} (\bibinfo {year}
  {2016})}\BibitemShut {NoStop}%
\bibitem [{\citenamefont {Williamson}\ \emph {et~al.}(2013)\citenamefont
  {Williamson}, \citenamefont {Schurmann}, \citenamefont {Hiller},
  \citenamefont {Klefenz}, \citenamefont {Hoerselmann}, \citenamefont {Husar},\
  and\ \citenamefont {Schober}}]{williamson2013}%
  \BibitemOpen
  \bibfield  {author} {\bibinfo {author} {\bibfnamefont {A.}~\bibnamefont
  {Williamson}}, \bibinfo {author} {\bibfnamefont {L.}~\bibnamefont
  {Schurmann}}, \bibinfo {author} {\bibfnamefont {L.}~\bibnamefont {Hiller}},
  \bibinfo {author} {\bibfnamefont {F.}~\bibnamefont {Klefenz}}, \bibinfo
  {author} {\bibfnamefont {I.}~\bibnamefont {Hoerselmann}}, \bibinfo {author}
  {\bibfnamefont {P.}~\bibnamefont {Husar}}, \ and\ \bibinfo {author}
  {\bibfnamefont {A.}~\bibnamefont {Schober}},\ }\bibfield  {title} {\enquote
  {\bibinfo {title} {Synaptic behavior and STDP of asymmetric nanoscale
  memristors in biohybrid systems},}\ }\href@noop {} {\bibfield  {journal}
  {\bibinfo  {journal} {Nanoscale}\ }\textbf {\bibinfo {volume} {5}},\ \bibinfo
  {pages} {7297--7303} (\bibinfo {year} {2013})}\BibitemShut {NoStop}%
\bibitem [{\citenamefont {Lindner}\ \emph {et~al.}(2004)\citenamefont
  {Lindner}, \citenamefont {Garc{\'\i}a-Ojalvo}, \citenamefont {Neiman},\ and\
  \citenamefont {Schimansky-Geier}}]{lindner2004}%
  \BibitemOpen
  \bibfield  {author} {\bibinfo {author} {\bibfnamefont {B.}~\bibnamefont
  {Lindner}}, \bibinfo {author} {\bibfnamefont {J.}~\bibnamefont
  {Garc{\'\i}a-Ojalvo}}, \bibinfo {author} {\bibfnamefont {A.}~\bibnamefont
  {Neiman}}, \ and\ \bibinfo {author} {\bibfnamefont {L.}~\bibnamefont
  {Schimansky-Geier}},\ }\bibfield  {title} {\enquote {\bibinfo {title}
  {Effects of noise in excitable systems},}\ }\href@noop {} {\bibfield
  {journal} {\bibinfo  {journal} {Phys. Rep.}\ }\textbf {\bibinfo {volume}
  {392}},\ \bibinfo {pages} {321--424} (\bibinfo {year} {2004})}\BibitemShut
  {NoStop}%
\bibitem [{\citenamefont {Pisarchik}\ and\ \citenamefont
  {Hramov}(2023)}]{pisarchik2023}%
  \BibitemOpen
  \bibfield  {author} {\bibinfo {author} {\bibfnamefont {A.}~\bibnamefont
  {Pisarchik}}\ and\ \bibinfo {author} {\bibfnamefont {A.}~\bibnamefont
  {Hramov}},\ }\bibfield  {title} {\enquote {\bibinfo {title} {Coherence
  resonance in neural networks: Theory and experiments},}\ }\href@noop {}
  {\bibfield  {journal} {\bibinfo  {journal} {Physics Reports}\ }\textbf
  {\bibinfo {volume} {1000}},\ \bibinfo {pages} {1--57} (\bibinfo {year}
  {2023})}\BibitemShut {NoStop}%
\bibitem [{\citenamefont {Nurzaman}\ \emph {et~al.}(2011)\citenamefont
  {Nurzaman}, \citenamefont {Matsumoto}, \citenamefont {Nakamura},
  \citenamefont {Shirai}, \citenamefont {Koizumi},\ and\ \citenamefont
  {Ishiguro}}]{nurzaman2011}%
  \BibitemOpen
  \bibfield  {author} {\bibinfo {author} {\bibfnamefont {S.}~\bibnamefont
  {Nurzaman}}, \bibinfo {author} {\bibfnamefont {Y.}~\bibnamefont {Matsumoto}},
  \bibinfo {author} {\bibfnamefont {Y.}~\bibnamefont {Nakamura}}, \bibinfo
  {author} {\bibfnamefont {K.}~\bibnamefont {Shirai}}, \bibinfo {author}
  {\bibfnamefont {S.}~\bibnamefont {Koizumi}}, \ and\ \bibinfo {author}
  {\bibfnamefont {H.}~\bibnamefont {Ishiguro}},\ }\bibfield  {title} {\enquote
  {\bibinfo {title} {From L{\'e}vy to Brownian: A computational model based on
  biological fluctuation},}\ }\href@noop {} {\bibfield  {journal} {\bibinfo
  {journal} {PLOS ONE}\ }\textbf {\bibinfo {volume} {6}},\ \bibinfo {pages}
  {1--11} (\bibinfo {year} {2011})}\BibitemShut {NoStop}%
\bibitem [{\citenamefont {Wu}, \citenamefont {Xu},\ and\ \citenamefont
  {Ma}(2017)}]{wu2017}%
  \BibitemOpen
  \bibfield  {author} {\bibinfo {author} {\bibfnamefont {J.}~\bibnamefont
  {Wu}}, \bibinfo {author} {\bibfnamefont {Y.}~\bibnamefont {Xu}}, \ and\
  \bibinfo {author} {\bibfnamefont {J.}~\bibnamefont {Ma}},\ }\bibfield
  {title} {\enquote {\bibinfo {title} {L{\'e}vy noise improves the electrical
  activity in a neuron under electromagnetic radiation},}\ }\href@noop {}
  {\bibfield  {journal} {\bibinfo  {journal} {PLOS ONE}\ }\textbf {\bibinfo
  {volume} {12}},\ \bibinfo {pages} {1--13} (\bibinfo {year}
  {2017})}\BibitemShut {NoStop}%
\bibitem [{\citenamefont {Chang}\ \emph {et~al.}(2011)\citenamefont {Chang},
  \citenamefont {Jo}, \citenamefont {Kim}, \citenamefont {Sheridan},
  \citenamefont {Gaba},\ and\ \citenamefont {Lu}}]{chang2011}%
  \BibitemOpen
  \bibfield  {author} {\bibinfo {author} {\bibfnamefont {T.}~\bibnamefont
  {Chang}}, \bibinfo {author} {\bibfnamefont {S.}~\bibnamefont {Jo}}, \bibinfo
  {author} {\bibfnamefont {K.}~\bibnamefont {Kim}}, \bibinfo {author}
  {\bibfnamefont {P.}~\bibnamefont {Sheridan}}, \bibinfo {author}
  {\bibfnamefont {S.}~\bibnamefont {Gaba}}, \ and\ \bibinfo {author}
  {\bibfnamefont {W.}~\bibnamefont {Lu}},\ }\bibfield  {title} {\enquote
  {\bibinfo {title} {Synaptic behaviors and modeling of a metal oxide
  memristive device},}\ }\href@noop {} {\bibfield  {journal} {\bibinfo
  {journal} {Applied Physics A}\ }\textbf {\bibinfo {volume} {102}},\ \bibinfo
  {pages} {857--863} (\bibinfo {year} {2011})}\BibitemShut {NoStop}%
\bibitem [{\citenamefont {Chen}\ \emph {et~al.}(2013)\citenamefont {Chen},
  \citenamefont {Li}, \citenamefont {Huang}, \citenamefont {Chen},
  \citenamefont {Wen},\ and\ \citenamefont {Qi}}]{chen2013}%
  \BibitemOpen
  \bibfield  {author} {\bibinfo {author} {\bibfnamefont {L.}~\bibnamefont
  {Chen}}, \bibinfo {author} {\bibfnamefont {C.}~\bibnamefont {Li}}, \bibinfo
  {author} {\bibfnamefont {T.}~\bibnamefont {Huang}}, \bibinfo {author}
  {\bibfnamefont {Y.}~\bibnamefont {Chen}}, \bibinfo {author} {\bibfnamefont
  {S.}~\bibnamefont {Wen}}, \ and\ \bibinfo {author} {\bibfnamefont
  {J.}~\bibnamefont {Qi}},\ }\bibfield  {title} {\enquote {\bibinfo {title} {A
  synapse memristor model with forgetting effect},}\ }\href@noop {} {\bibfield
  {journal} {\bibinfo  {journal} {Phys. Lett. A}\ }\textbf {\bibinfo {volume}
  {377}},\ \bibinfo {pages} {3260--3265} (\bibinfo {year} {2013})}\BibitemShut
  {NoStop}%
\bibitem [{\citenamefont {Zhou}, \citenamefont {Fang},\ and\ \citenamefont
  {Yang}(2019)}]{zhou2019}%
  \BibitemOpen
  \bibfield  {author} {\bibinfo {author} {\bibfnamefont {E.}~\bibnamefont
  {Zhou}}, \bibinfo {author} {\bibfnamefont {L.}~\bibnamefont {Fang}}, \ and\
  \bibinfo {author} {\bibfnamefont {B.}~\bibnamefont {Yang}},\ }\bibfield
  {title} {\enquote {\bibinfo {title} {A general method to describe forgetting
  effect of memristors},}\ }\href@noop {} {\bibfield  {journal} {\bibinfo
  {journal} {Phys. Lett. A}\ }\textbf {\bibinfo {volume} {383}},\ \bibinfo
  {pages} {942--948} (\bibinfo {year} {2019})}\BibitemShut {NoStop}%
\bibitem [{\citenamefont {Janicki}\ and\ \citenamefont
  {Weron}(1994)}]{janicki1994}%
  \BibitemOpen
  \bibfield  {author} {\bibinfo {author} {\bibfnamefont {A.}~\bibnamefont
  {Janicki}}\ and\ \bibinfo {author} {\bibfnamefont {A.}~\bibnamefont
  {Weron}},\ }\href@noop {} {\emph {\bibinfo {title} {Simulation and Chaotic
  Behavior of Alpha-stable Stochastic Processes}}}\ (\bibinfo  {publisher}
  {Published by Marcel Dekker, New York},\ \bibinfo {year} {1994})\BibitemShut
  {NoStop}%
\bibitem [{\citenamefont {Weron}\ and\ \citenamefont
  {Weron}(1995)}]{weron1995}%
  \BibitemOpen
  \bibfield  {author} {\bibinfo {author} {\bibfnamefont {A.}~\bibnamefont
  {Weron}}\ and\ \bibinfo {author} {\bibfnamefont {R.}~\bibnamefont {Weron}},\
  }\enquote {\bibinfo {title} {Computer simulation of L{\'e}vy $\alpha$-stable
  variables and processes},}\ \ (\bibinfo  {publisher} {Springer},\ \bibinfo
  {year} {1995})\ pp.\ \bibinfo {pages} {379--392}\BibitemShut {NoStop}%
\bibitem [{\citenamefont {Korneev}, \citenamefont {Zakharova},\ and\
  \citenamefont {Semenov}(2024)}]{korneev2024}%
  \BibitemOpen
  \bibfield  {author} {\bibinfo {author} {\bibfnamefont {I.}~\bibnamefont
  {Korneev}}, \bibinfo {author} {\bibfnamefont {A.}~\bibnamefont {Zakharova}},
  \ and\ \bibinfo {author} {\bibfnamefont {V.}~\bibnamefont {Semenov}},\
  }\bibfield  {title} {\enquote {\bibinfo {title} {L{\'e}vy noise-induced
  coherence resonance: numerical study versus experiment},}\ }\href@noop {}
  {\bibfield  {journal} {\bibinfo  {journal} {Chaos, Solitons and Fractals}\
  }\textbf {\bibinfo {volume} {184}},\ \bibinfo {pages} {115037} (\bibinfo
  {year} {2024})}\BibitemShut {NoStop}%
\bibitem [{\citenamefont {Mannella}(2002)}]{mannella2002}%
  \BibitemOpen
  \bibfield  {author} {\bibinfo {author} {\bibfnamefont {R.}~\bibnamefont
  {Mannella}},\ }\bibfield  {title} {\enquote {\bibinfo {title} {Integration of
  stochastic differential equations on a computer},}\ }\href@noop {} {\bibfield
   {journal} {\bibinfo  {journal} {International Journal of Modern Physics C}\
  }\textbf {\bibinfo {volume} {13}},\ \bibinfo {pages} {1177--1194} (\bibinfo
  {year} {2002})}\BibitemShut {NoStop}%
\bibitem [{\citenamefont {Xu}\ \emph {et~al.}(2016)\citenamefont {Xu},
  \citenamefont {Li}, \citenamefont {Zhang}, \citenamefont {Li},\ and\
  \citenamefont {Kurths}}]{xu2016}%
  \BibitemOpen
  \bibfield  {author} {\bibinfo {author} {\bibfnamefont {Y.}~\bibnamefont
  {Xu}}, \bibinfo {author} {\bibfnamefont {Y.}~\bibnamefont {Li}}, \bibinfo
  {author} {\bibfnamefont {H.}~\bibnamefont {Zhang}}, \bibinfo {author}
  {\bibfnamefont {X.}~\bibnamefont {Li}}, \ and\ \bibinfo {author}
  {\bibfnamefont {J.}~\bibnamefont {Kurths}},\ }\bibfield  {title} {\enquote
  {\bibinfo {title} {The switch in a genetic toggle system with L{\'e}vy
  noise},}\ }\href@noop {} {\bibfield  {journal} {\bibinfo  {journal}
  {Scientific Reports}\ }\textbf {\bibinfo {volume} {6}},\ \bibinfo {pages}
  {31505} (\bibinfo {year} {2016})}\BibitemShut {NoStop}%
\bibitem [{\citenamefont {Pavlyukevich}\ \emph {et~al.}(2010)\citenamefont
  {Pavlyukevich}, \citenamefont {Dybiec}, \citenamefont {Chechkin},\ and\
  \citenamefont {Sokolov}}]{pavlyukevich2010}%
  \BibitemOpen
  \bibfield  {author} {\bibinfo {author} {\bibfnamefont {I.}~\bibnamefont
  {Pavlyukevich}}, \bibinfo {author} {\bibfnamefont {B.}~\bibnamefont
  {Dybiec}}, \bibinfo {author} {\bibfnamefont {A.}~\bibnamefont {Chechkin}}, \
  and\ \bibinfo {author} {\bibfnamefont {I.}~\bibnamefont {Sokolov}},\
  }\bibfield  {title} {\enquote {\bibinfo {title} {L{\'e}vy ratchet in a weak
  noise limit: Theory and simulation},}\ }\href@noop {} {\bibfield  {journal}
  {\bibinfo  {journal} {Eur. Phys. J. Special Topics}\ }\textbf {\bibinfo
  {volume} {191}},\ \bibinfo {pages} {223--237} (\bibinfo {year}
  {2010})}\BibitemShut {NoStop}%
\bibitem [{Note1()}]{Note1}%
  \BibitemOpen
  \bibinfo {note} {We continuously vary the coupling strengths with certain
  parameter step such that the last state at the previous iteration is the
  initial condition for the next one.}\BibitemShut {Stop}%
\end{thebibliography}


\end{document}